# Experimental observation of non-Abelian earring nodal links in phononic crystals

Mudi Wang[1], Shan Liu[2], Qiyun Ma[2], Ruo-Yang Zhang[1*], Dongyang Wang[1], Qinghua Guo[1,3], Biao Yang[1,4], Manzhu Ke[2], Zhengyou Liu[2,5*], and C. T. Chan[1*]

[1]Department of Physics, The Hong Kong University of Science and Technology, Hong Kong, China

[2]Key Laboratory of Artificial Micro- and Nanostructures of Ministry of Education and School of Physics and Technology, Wuhan University, Wuhan, China

[3]Institute for Advanced Study, The Hong Kong University of Science and Technology, Hong Kong, China

[4]College of Advanced Interdisciplinary Studies, National University of Defense Technology, Changsha 410073, China

[5]Institute for Advanced Studies, Wuhan University, Wuhan, China

*Correspondence to: zhangruoyang@gmail.com; zyliu@whu.edu.cn; phchan@ust.hk

**Nodal lines are symmetry-protected one-dimensional band degeneracies in momentum space, which can appear in numerous topological configurations such as nodal rings, chains, links, and knots. Very recently, non-Abelian topological physics has been proposed in space-time inversion ($PT$) symmetric systems, and attract widespread attention. One of the most special configurations in non-Abelian system is the earring nodal link, composing of a nodal chain linking with an isolated nodal line, is signature of non-Abelian topology and cannot be elucidated using Abelian topological classifications. However, the earring nodal links have not been yet observed in real system. Here we design the phononic crystals with earring nodal links, and verify its non-Abelian topologicial charge in full-wave simulations. Moreover, we experimentally observed two different kinds of earring nodal links by measuring the band structures for two phononic crystals. Specifically, we found that the order of the nodal chain and line can switch after band inversion but their link cannot be severed. Our work provides experimental evidence for phenomena unique to non-Abelian band topology and our simple acoustic system provides a convenient platform for studying non-Abelian charges.**

A main theme in fundamental physics and materials science is the discovery of new topological phases. Topological materials, such as topological insulators[1-11], Weyl or Dirac semimetals[12-19] and nodal line semimetals[20-24], have attracted much attention in both theory and experiment. All of them can be characterized by quantized topological numbers and can be classified by Abelian groups $Z$ or $Z_2$. Very recently, a new type of topological classification using non-Abelian groups has been proposed for space-time inversion (*PT*) symmetric multiband systems[25], promoting the discovery of many braiding topological structures[26-35] in non-Abelian systems.

The band structure in non-Abelian systems supports some special configurations in moment space protected by non-Abelian topological charges. One of the most interesting configurations is the earring nodal link[25], the existence and robustness of which cannot be explained by conventional Abelian topology that defines topological invariants by considering one single bandgap. Such earring nodal links can be observed in a 3-band system, in which the non-Abelian topology can be described by a quaternion charge of "-1". In this work, we report the first experimental observation of earring nodal links and their non-trivial evolution as system parameter changes. We designed and fabricated three-dimensional (3D) phononic crystals with three-band dispersion and we experimentally observed two kinds of earring nodal links.

In a three-band *PT*-symmetric system, the space of the Hamiltonian for a loop which does not touch any nodal lines in the momentum space is $M_3 = O(3)/O(1)^3$, where $O(N)$ is the orthogonal group and $O(1) \equiv \pm 1$. Its fundamental homotopy group is the non-Abelian quaternion group $\pi_1(M_3) = Q = \{-1, \pm i, \pm j, \pm k, +1\}$, which satisfies $i^2 = j^2 = k^2 = -1$, and $ij = -ji = k$. The characteristics of global nodal line configuration can be described by these non-Abelian topological charges. In this work, we will focus on the quaternion charge -1.

Let us consider a 3-band 3D system and we start with constructing a triple degeneracy lying on the intersection of two mirror-invariant planes in the momentum space. The little group of a ***k*** point on the $k_z$ axis (i.e. the intersection of two mirror-invariant planes $k_x = 0$ and $k_y = 0$) is $C_{2v}$. Suppose that two out of the three eigenmodes have both opposite parities on the *x* and *y* mirror and mirror planes on the $k_z$ axis, we can write down the 3-by-3 representations of the group elements as

$$M_x = \begin{pmatrix} -1 & 0 & 0 \\ 0 & 1 & 0 \\ 0 & 0 & 1 \end{pmatrix}, M_y = \begin{pmatrix} 1 & 0 & 0 \\ 0 & -1 & 0 \\ 0 & 0 & 1 \end{pmatrix}. \tag{1}$$

Then, (see supplementary materials) the general form of the $k \cdot p$ Hamiltonian respecting $PT$ and $C_{2v}$ symmetries near the $k_z$ axis (up to second order of $k_x, k_y$) can be written as

$$H(k_x, k_y, k_z) = \begin{pmatrix} f_1(k_z, k_x^2, k_y^2) & w_{xy}k_xk_y & (v_x^0 + v_x^1 k_z)k_x \\ w_{xy}k_xk_y & f_2(k_z, k_x^2, k_y^2) & (v_y^0 + v_y^1 k_z)k_y \\ (v_x^0 + v_x^1 k_z)k_x & (v_y^0 + v_y^1 k_z)k_y & f_3(k_z, k_x^2, k_y^2) \end{pmatrix}, \tag{2}$$

where $f_i(k_z, k_x^2, k_y^2) = g_i(k_z) + \sum_{j=x,y} k_j^2 h_{ij}(k_z)$ ($j = 1,2,3$). To obtain a three-fold degeneracy along the $k_z$ axis, we require $g_1(k_z) = g_2(k_z) = g_3(k_z)$. Since these identities cannot be satisfied simultaneously at a general $k_z$ point, a triple degeneracy point does not always exist on the $k_z$ axis. However, such triple degeneracy can be accidentally achieved at a point for some specific system parameters. As an example, we consider a specific Hamiltonian

$$H_1(k_x, k_y, k_z) = \begin{pmatrix} k_z & 0.5k_xk_y & 1.5\,k_x \\ 0.5k_xk_y & 0.3k_z + b & (1 - 0.2k_z)\,k_y \\ 1.5\,k_x & (1 - 0.2k_z)\,k_y & 0 \end{pmatrix}, \tag{3}$$

which is special case of $H$ in Eq. (2). The distribution of nodal lines of this $H_1$ is shown in Fig. 1a. There is a triple point is at $k_z = 0$ when $b = 0$, and the Hamiltonian (up to first order and a frequency shift) in the transverse plane will reduce to

$$H_1(k_x, k_y, k_z = 0) = v_x^0\, k_x \hat{\lambda}_2 + v_y^0 k_y \hat{\lambda}_3$$
$$= \begin{pmatrix} 0 & 0 & v_x^0\, k_x \\ 0 & 0 & v_y^0\, k_y \\ v_x^0\, k_x & v_y^0\, k_y & 0 \end{pmatrix}. \tag{4}$$

Here, $\hat{\lambda}_i$ denote the Gell-Mann matrices, and $\hat{\lambda}_2, \hat{\lambda}_3, \hat{\lambda}_4$ satisfy the angular momentum commutation relation $[\hat{\lambda}_i, \hat{\lambda}_j] = i\epsilon_{ijk}\hat{\lambda}_k$ ($i, j = 2,3,4, i \neq j$). As such, they form a representation of spin-1 operators $\hat{\lambda}_2 = \hat{S}_x$, $\hat{\lambda}_3 = \hat{S}_y$, $\hat{\lambda}_4 = \hat{S}_z$. The transverse Hamiltonian at the accidental triple point, with band dispersion shown in Fig. 1b, is in fact the 2-dimensional (2D) spin-1 Hamiltonian of a Dirac-like cone[36-47] that has been studied widely as an effectively zero refractive index system: $H_1(k_x, k_y, k_z = 0) = v_x^0\, k_x \hat{S}_x + v_y^0 k_y \hat{S}_y$. This can be rewritten as $H_1(k_x, k_y) = H_1(\tilde{k}, \phi) = \tilde{k} \exp(i\,\hat{S}_z \phi)\,\hat{S}_x$

$\exp(-i\,\hat{S}_z\phi) = \tilde{k}\exp(\hat{L}_z\phi)\hat{S}_x\exp(-\hat{L}_z\phi)$, where $\tilde{k}e^{i\phi} = v_x^0 k_x + iv_y^0 k_y$ gives the generalized amplitude and argument of the wave vector, and $\hat{L}_z = i\hat{S}_z = \begin{pmatrix} 0 & 1 & 0 \\ -1 & 0 & 0 \\ 0 & 0 & 0 \end{pmatrix}$ denotes the generator of rotations about the z axis, and $\hat{S}_x = H_1(1,0)$ gives the base point of the Hamiltonian on the unit circle. When the wave vector winds around the triple point one time along the green loop in Fig. 1a, the generalized argument also changes $2\pi$. As a result, the triple point behaves as a topological defect of the orthonormal frame formed by the eigenvectors of the three bands, around which the frame of eigenvectors rotates one turn about the z axis as shown by the distribution of eigenvectors in Fig. 1c. The three color bars at each point in Fig. 1c denote the three orthogonal real-valued eigenvectors at that point. Therefore, we have demonstrated that the spin-1 Hamiltonian corresponds to $2\pi$ rotation of all three eigenstates about the fixed axis (z axis), and such eigenstate frame rotation is the key feature of the non-Abelian quaternion charge -1[25], as shown in the Fig. 1d which shows the rotation of the eigenstates frame on the unit sphere along green loop in Fig. 1a. From the Abelian topological viewpoint, the 2D spin-1 triple point in Fig. 1b is entirely accidental and has no topological protection because the Berry phase around the triple point of every band is zero. Our results, however, show that the 2D spin-1 triple point is topologically nontrivial, and protected by the quaternion topological charge −1 characterized by the winding of eigenstates around the triple point. This explains why the degeneracies between the three bands forming the Dirac-like cone in 2D PT symmetric crystals can never be fully gapped and their evolution is governed by the non-Abelian charge.

The above analysis indicates that this triple degenerate point as the crossing point of two nodal lines in Fig. 1a, is protected by non-Abelian quaternion charges −1. This is in line with the non-Abelian topology analysis of multiple band systems proposed recently[25,29]: the green loop encircles two nodal lines with the same orientation formed by the same pair of bands (see supplementary text). These two nodal lines have to be linked together in some ways and cannot be completely separated by tuning system parameters. As shown in Fig. 1e, when b is changed to 0.15, a new nodal ring (red circle) of the lower two bands appears between two nodal lines, which forms a nodal link with the blue nodal line, and forms a nodal chain with the red nodal line. The underlying mechanism is that the non-trivial non-Abelian quaternion charge −1 cannot be changed by small perturbations, so the green loop in Fig. 1e must encircle at least two nodal lines with the same orientation of the same pair of bands, which ensures the existence of a nodal ring between the two nodal lines. When a different perturbation is added, such as $b$ is changed to −0.15, another configuration of earring nodal link appears, as shown in Fig. 1f. A new emerging blue nodal ring of the upper two bands and the blue line form a nodal chain, while the blue nodal ring and red nodal line form a nodal link. These hybrid configurations of a nodal link and nodal chain, which are called earring nodal links[25], are the special nodal structures in non-Abelian topology and cannot be explained by the Abelian topology.

In order to observe experimentally earring nodal links similar to those shown in Fig. 1e and Fig. 1f, we designed a cubic layer-by-layer phononic crystal with a unit cell consisting of two stacked cuboid that are twisted by $\pi/2$ along the z direction. The unit cell is shown in Fig. 2a and the structural parameters are: $a = 22.6$mm, $b = 18.8$mm, $w_2 = 7.1$mm, $h_1 = 6$mm, $h_2 = 6$mm. Our full-wave simulations are performed using COMSOL Multiphysics. As the structure has mirror symmetries in both the $x$ and $y$ directions, each mode has well defined mirror parity in the corresponding mirror planes. In Fig. 2b, (+, +) indicates that the sound pressure field of the eigenmode is even in both $x$ and $y$ directions on the intersection of two mirror planes $k_x = k_y = 0$. While (+, -) indicates that the pressure field is even mode in $x$ and odd mode in $y$ direction, and (-, +) indicates that the pressure field is odd mode in $x$ and even mode in $y$ direction. The simulated dispersion curves along $k_z$ direction with $w_1 = 1.6$mm are shown in Fig. 2b, and the pressure field distributions for the eigenstates in bands 2-4 have three different mode symmetries (+, +), (+, -) and (-, +) as indicated in the figure. The three bands intersect each other, and the red (blue) points represent the nodal point of the two lower (higher) bands. When $w_1$ is increased to 3.0mm, the bands 2-4 become accidentally degenerate at one point, as shown in Fig. 2c. Continuing to increase $w_1$ to 3.9mm, bands 2-4 will intersect each other again. The nodal lines in $k$ space are found numerically and are shown in Fig. 2e-g (solid lines). They bear the same features as the model Hamiltionian shown in Fig. 1e,a,f. In order to verify the non-Abelian topological charge in full-wave simulations, we extract the cell-periodic part of Bloch eigenfields $u_k(r) =$

$e^{-i\mathbf{k}\cdot\mathbf{r}}p_{\mathbf{k}}(\mathbf{r})$ from the Bloch sound pressure fields $p_{\mathbf{k}}(\mathbf{r})$ along the green loop such that they satisfy the *PT* symmetric gauge $u_{\mathbf{k}}(\mathbf{r}) = u_{\mathbf{k}}(-\mathbf{r})^*$. Fixing a basepoint $\mathbf{k}_0$ on the loop, the three eigenfields $u_{\mathbf{k}_0}^i(\mathbf{r})$ ($i = 1,2,3$) at $\mathbf{k}_0$ form a basis. Then we can obtain three effective 3-component eigenvectors $\vec{v}^i(\mathbf{k})$ ($i = 1,2,3$) via projecting the eigenfields at each point $\mathbf{k}$ on the loop onto the basis $[\vec{v}^i(\mathbf{k})]_j = \langle u_{\mathbf{k}_0}^j(\mathbf{r}) | u_{\mathbf{k}}^i(\mathbf{r}) \rangle$. It can be shown that the 3-component vectors $\vec{v}^i(\mathbf{k})$ are guaranteed to be orthogonal real vectors by the *PT* symmetric gauge. The obtained effective eigenstate frames along the loop are plotted on the unit sphere in Fig. 2h. It shows the $2\pi$ rotation of all three eigenstates on the unit sphere about the z axis, which confirms that the earring nodal link in the real phononic crystal is protected by a quaternion charge $-1$.

Using 3D printing, we fabricated an experimental sample containing $21 \times 21 \times 21$ unit cells, as shown in Fig. 3a. The sample has the same symmetry as the Hamiltonian $H_1$ and has the earring nodal links as shown in Fig. 2e. We place an acoustic point source at the position of the red star. We insert a movable microphone (diameter ~0.7 cm, B&K Type 4187) into the sample through the interstitial voids to scan the acoustic field of the whole sample. The band dispersions on the planes of $k_x = 0$ and of $k_y = 0$ at different frequencies are obtained experimentally by doing 3D Fourier transforms of the scan fields, and are compared with the nodal lines in theory (white lines), as shown in Fig. 3b-d. At each frequency, four theoretical nodal points are marked by red or green colors, helping us to find the nodal points from the experimental data which are located at the intersections of equal-frequency contours. The nodal points in Fig. 3b correspond to the 4 red dots in Fig. 2e at $k_y = 0$ plane. While the nodal points in Fig. 3c,d correspond to the red and blue dots in Fig. 2e at the $k_x = 0$ plane respectively. Hence the nodal links with earring shape are verified experimentally.

In order to verify the exotic evolution of the earing nodal link protected by non-Abelian $-1$ charge, another sample with $w_1 = 3.9$mm corresponding to Fig. 2g was fabricated. The band dispersions obtained using the Fourier transform of scanned fields are shown in Fig. 3f-h for different frequencies, and we can see the position of nodal point with the help of theoretical result. Hence the other configuration of earring nodal links, whose existence is protected by quaternion charge -1, is also observed in experiment.

We proved theoretically that the 2-dimensional spin-1 triple point is indeed protected by the non-Abelian quaternion charge -1, which shows that the horizon of topological protection can be broadened if we extend the scope from Abelian to non-Abelian. We also revealed that indestructible earring nodal links can emerge from the perturbation of the triple point, and numerically verified the non-Abelian topological charge -1 carried by nodal link from full-wave simulations. Moreover, we experimentally observed the earring nodal links that were predicted to exist in non-Abelian systems, and the earring nodal links before and after band inversion are measured at two system configurations, providing evidence for the nontrivial evolution of the earing nodal link and manifests experimentally the stability of non-Abelian topological quaternion charge -1. Our results provide a solid experimental basis for the theory of non-Abelian band topology and offer a simple three-dimensional acoustic system as a platform to explore new phenomena associated with mutlti-band topology.

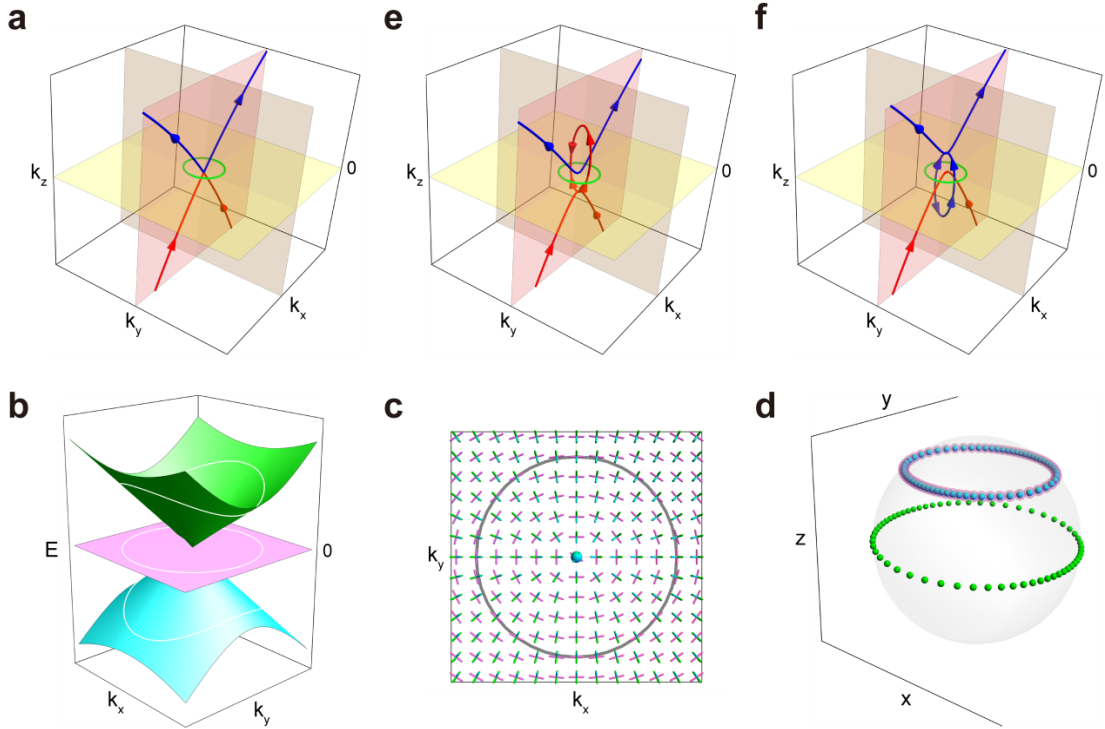

**Fig. 1. Nontrivial accidental triple degeneracy in a non-Abelian system and its evolution. a,** The nontrivial accidental triple degeneracy in momentum space. The red (blue) nodal lines are formed by the lower (upper) pair of bands. **b,** The dispersion at $k_z = 0$. **c,** Eigenstates on the $k_z = 0$ plane, where the three color bars represent the three orthogonal eigenstates, and the gray circle corresponds to the green loop in Fig. 1a. **d,** The rotation of the eigenstates frame on the unit sphere along green loop in Fig. 1a, where the spheres with the same color trace out the trajectory of the eigenstates of each band. In **b-d,** cyan, lighter magenta and green colors correspond to bands 1, 2, and 3, respectively. **e,f,** The two earring nodal links configurations when two different perturbations (changing the value of $b$ from 0 to 0.15 and -0.15 respectively) are added to the case of the nontrivial accidental triple degeneracy.

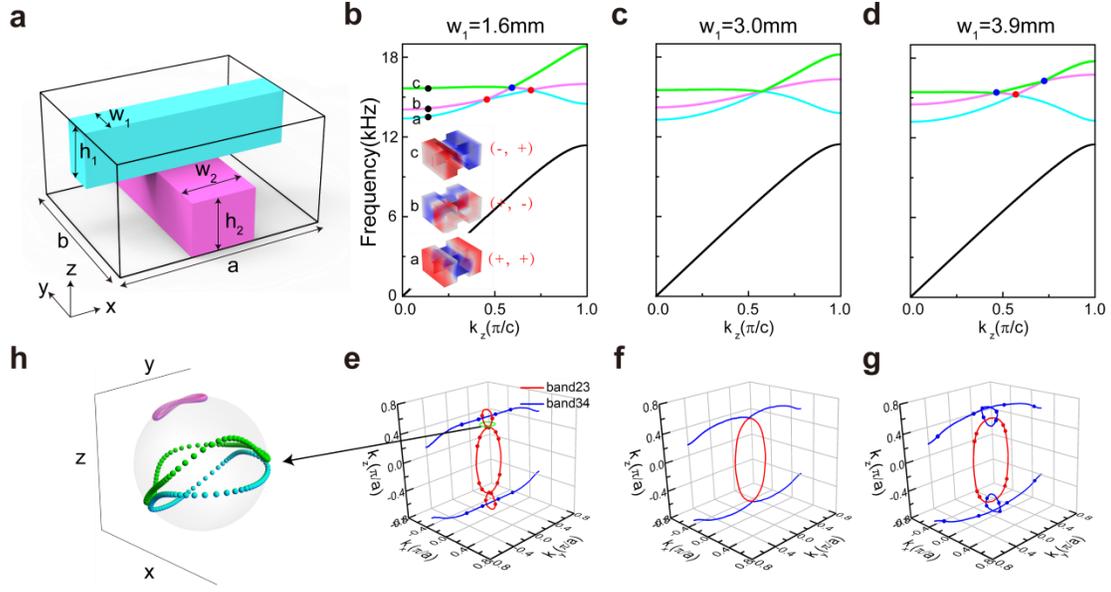

**Fig. 2. The evolution of band structure in the non-Abelian system. a,** The unit cell of the phononic crystals. **b-d,** The corresponding band structures along $k_z$ with $w_1$= 1.6, 3.0 and 3.9 mm. The black dots mark three eigenstates in bands 2-4 with different symmetries, and the distribution of the pressure fields are shown in the illustration. The red (blue) dots represent the nodal points of band 23 (band 34). **e-g,** Distribution of nodal lines in **e-g** corresponding to the band structures in **b-d**, respectively. The red and blue dots in **e** (**g**) correspond to the red and blue nodal points in Fig. 3b-d (Fig. 3f-h). **h,** The rotation of the eigenstates frame on the unit sphere along green loop in full-wave simulations.

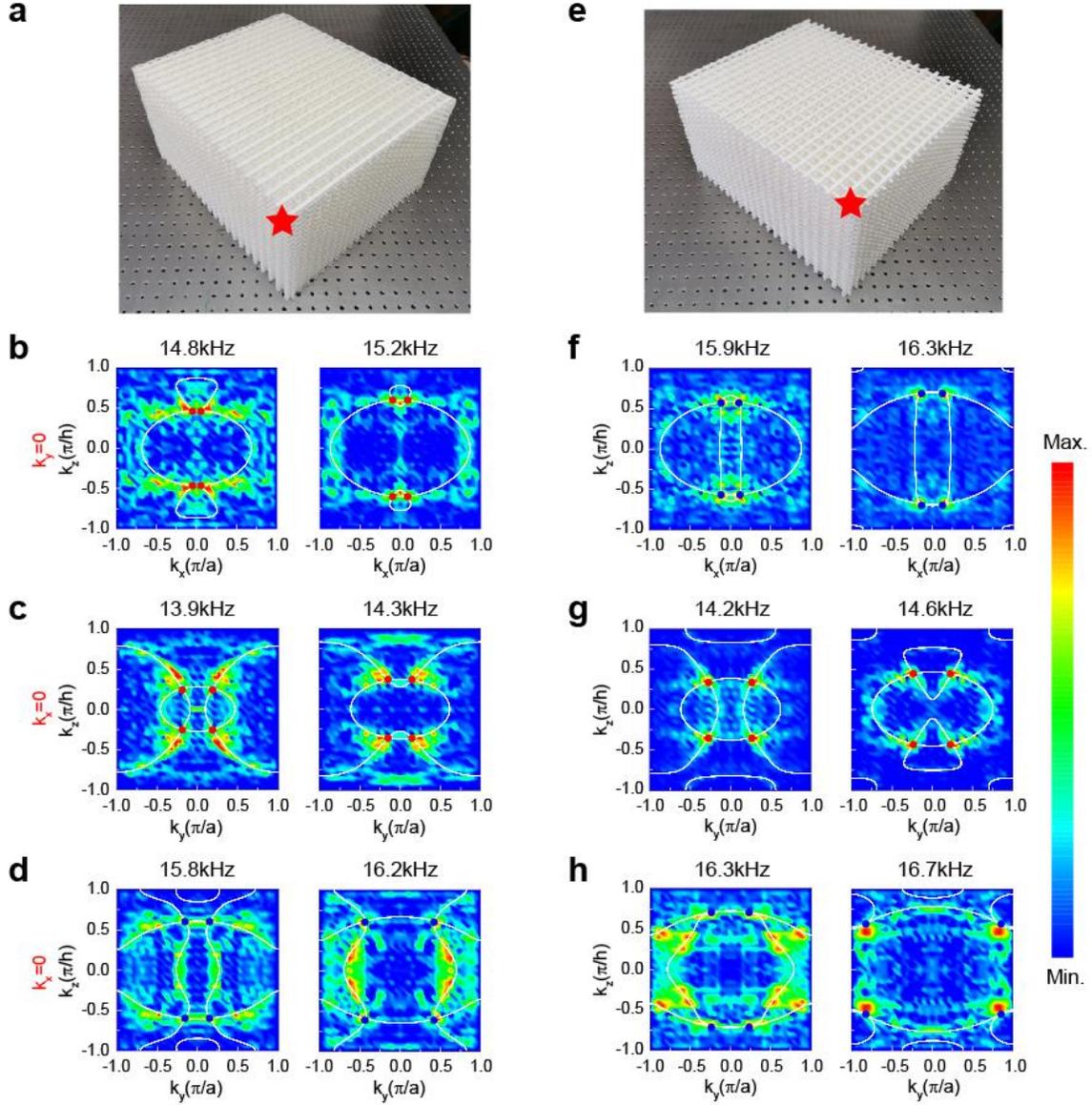

**Fig. 3. Experimental observation of two configurations of earring nodal links protected by the non-trivial non-Abelian quaternion charge -1. a,** The experimental sample-1 (the structural parameters are the same as Fig. 2e). The red star is the position of the point source. **b-d,** The experimental (color maps) and theoretical (white lines) equal-frequency contours at $k_y=0$ and $k_x=0$ for different frequencies for sample-1. The nodal points in theory are marked by red or green dots. **e,** The experimental sample-2 (the structural parameters are the same as Fig. 2g). **f-g,** The experimental and theoretical band structures at $k_y=0$ and $k_x=0$ for different frequencies for sample-2. The nodal points in the theory are marked by red or blue dots.

**Acknowledgment**

This work is supported by Hong Kong RGC grant AoE/P-02/12 and KAUST CRG grant (KAUST20SC01) and by the Croucher foundation (CAS20SC01). Z. L. is supported by the National Natural Science Foundation of China (Grants No. 11890701, 11774275) and the National Key R&D Program of China (Grant No. 2018YFA0305800).

**Author contributions**

M.W., R.-Y.Z., Z. L., and C.T.C. conceived the idea. M.W. did the simulations and designed the experimental samples. M.W., S. L., Q. M., and M. K carried out all the measurements. M.W., R.-Y.Z., D. W., Q. G. did the data analysis. R.-Y.Z., Z. L., and C.T.C. supervised the whole project. M.W. wrote the draft. R.-Y.Z., B.Y., Z.L and C.T.C. analyzed the data and revised the manuscript.